\documentclass[reprint,onecolumn,amsmath,amssymb,aps,nofootinbib,pre]{revtex4-2}

\usepackage{graphicx} 
\usepackage[colorlinks=true,linkcolor=blue,citecolor=blue,urlcolor=blue]{hyperref}
\usepackage{bm}
\usepackage{comment}
\allowdisplaybreaks

\begin{document}

\title{Geometric Propulsion of a Scallop with Control-Dependent Hydrodynamics}
\author{Hisao Hayakawa}
\email[e-mail: ]{hisao@yukawa.kyoto-u.ac.jp}
\affiliation{Center for Gravitational Physics and Quantum Information, Yukawa Institute for Theoretical Physics, Kyoto University, Kitashirakawa-Oiwakecho, Sakyo-ku, Kyoto 606-8502, Japan}

\date{\today}

\begin{abstract}

We study a force-free single-hinge swimmer driven by two controls.
When its axial hydrodynamic connection depends only on the opening angle, we show that the mean displacement per cycle vanishes in the periodic state at any driving rate. 
An independent control of the hydrodynamic response removes this restriction. For a factorized connection,
the displacement at arbitrary driving rate is an integral of a measurable shape moment against the hydrodynamic modulation; in the slow-driving limit it reduces to a curvature flux which, at fixed potential and connection, is independent of the shape mobility.
A reflecting harmonic model illustrates how thermal fluctuations modify this flux without generating it, and a hinged-bead model with variable passive slip gives a quantitative opening--closing test of the mechanism. We distinguish these propulsion results from the leading shape-sector dissipation, which depends on the mobility and is governed by a thermodynamic metric.

\end{abstract}

\maketitle

\section{Introduction and Theoretical Framework}

Locomotion at low Reynolds numbers (\(Re \ll 1\)) is governed by the linear, time-independent Stokes equations, where viscous drag dominates over inertial forces \cite{Purcell1977,HappelBrenner}. In 1977, Purcell formulated the \emph{scallop theorem}, stating that a microswimmer executing a periodic, time-reversible sequence of shape deformations, such as a single-hinge scallop opening and closing its shell, cannot achieve net displacement over a full cycle in an unbounded Newtonian fluid \cite{Purcell1977,Lauga2009}.

From a geometric viewpoint, Shapere and Wilczek \cite{Shapere1989} reformulated microswimming in terms of a gauge connection over shape space. Building upon this idea, Ishimoto and Yamada \cite{Ishimoto2012} developed a coordinate-free formulation of low-Reynolds-number locomotion, in which the translation generated by a closed deformation cycle is related to the geometric structure of the corresponding shape space. In this framework, Purcell's scallop theorem has a particularly simple interpretation: a one-degree-of-freedom (1-DOF) shape manifold is one-dimensional, and a reciprocal cycle cannot enclose a finite area in shape space. Deterministic microswimmers therefore generally require at least two independent shape degrees of freedom to achieve geometric propulsion, as exemplified by Purcell's three-link swimmer \cite{Purcell1977} and the Najafi--Golestanian three-sphere swimmer \cite{Najafi2004}.

At microscopic scales, however, internal shape variables are inevitably affected by thermal fluctuations and other stochastic forces \cite{Bregulla2014}. An internal coordinate \(x\), such as the opening angle of a single-hinge scallop, should then be regarded as a stochastic variable described by a probability density \(\rho(x,t)\) obeying an overdamped Fokker--Planck or Smoluchowski equation \cite{Lau2009,Golestanian2008,Astumian2002}. This raises a natural question: can stochastic shape dynamics driven by multiple external controls circumvent the geometric restriction underlying the scallop theorem?

An important observation of the present work is that the answer is negative if the hydrodynamic response depends only on the instantaneous shape. For a force-free 1-DOF swimmer whose translational velocity can be written as
\(\dot X=-G(x)\dot x ,\)
the infinitesimal displacement 
\begin{align}\label{eq(1.1)}
    dX=-G(x)dx
\end{align}
is an exact differential. 
Consequently, its integral over any closed trajectory vanishes, irrespective of whether \(x\) evolves deterministically or stochastically. Thus, thermal fluctuations and a two-dimensional control protocol alone are insufficient to generate propulsion. The conventional scallop theorem survives the introduction of stochastic shape dynamics.

This observation identifies the additional ingredient required for geometric propulsion: the hydrodynamic connection itself must depend on an independently controlled parameter. We therefore consider a more general force-free relation
\begin{align}
\dot X=-G(x,\boldsymbol{\theta})\dot x,\label{eq:controlled_kinematics}
\end{align}
where \(\boldsymbol{\theta}=(\theta_1,\theta_2)\) denotes two externally controlled parameters. Physically, such control dependence may arise when the external driving modifies not only the internal potential governing the shape dynamics but also a hydrodynamic property of the swimmer, for example through a controllable surface state, boundary condition, or mechanical property coupled to the fluid. As a minimal realization, we introduce a weak control dependence into the angle-dependent hydrodynamic coupling \(G_{\rm Kim}(x)\)~\cite{Kim1986},
\begin{align}\label{G(x,theta_2)}
G(x,\theta_2)
=
G_{\rm Kim}(x)
\left[
1+\epsilon
\left(
\frac{\theta_2}{\overline{\theta}_2}-1
\right)
\right],
\end{align}
where \(G_{\rm Kim}(x)\) describes the angle dependence of the hydrodynamic coupling~\cite{Kim1986}, and \(\epsilon\) measures the strength of the hydrodynamic modulation. This phenomenological form may be regarded as the leading-order expansion of a more general control-dependent hydrodynamic connection around a reference value of \(\theta_2\).

The resulting mechanism is naturally related to geometric pumping in stochastic systems. Cyclic modulation of control parameters can generate geometric contributions to integrated currents associated with Berry-like phases, often referred to as Berry--Sinitsyn--Nemenman (BSN) phases \cite{thouless1,thouless2,berry,xiao,sinitsyn1,sinitsyn2}. Such geometric pumping has been experimentally demonstrated in charge transport \cite{ex-ch1,ex-ch2,ex-ch2.5,ex-ch3,ex-ch5,ex-thou1,ex-thou2} and spin pumping \cite{ex-spin1}, and has been theoretically studied using scattering approaches \cite{brouwer,s-th-ch1,s-th-ch2,s-th-ch3,s-th-spin1}, classical master equations \cite{parrondo,ren,sagawa,ville2021}, and quantum master equations \cite{qme1,yuge1}. The same geometric viewpoint has also been extended to thermodynamic settings, leading to geometric heat engines and adiabatic pumps driven by cyclic modulation of reservoir parameters and Hamiltonian couplings \cite{Wang2021,Brandner-Saito,engine,Hino2021,Bhandari20,Abiuso20,Alonso21,Eglinton,Yoshii22,Wang2024,Breuer2016,FHHT,Funo2020,Yoshii2026}.

In this paper, we combine these ideas to formulate a stochastic 1-DOF scallop with control-dependent hydrodynamics. We show that the slow cyclic modulation of two control parameters generates a geometric curvature in control space whose flux determines the net displacement per cycle. The conventional scallop theorem is continuously recovered when the control dependence of the hydrodynamic connection is removed. For sufficiently small driving cycles, the displacement obeys a geometric area law. Thermal fluctuations are not the origin of propulsion in the present model; rather, they modify the stochastic shape distribution and suppress the geometric displacement at high temperature. In parallel, the nonadiabatic excess dissipation is characterized by a Riemannian thermodynamic metric, allowing propulsion and energetic cost to be described within a common geometric framework.

We emphasize that the scallop considered here is an idealized low-Reynolds-number swimmer rather than a model of the propulsion mechanism of an actual scallop, which can employ fluid jets and operate outside the assumptions underlying the Stokes description. The purpose of the model is instead to identify the minimal geometric and hydrodynamic ingredients required for a fluctuating single-shape-coordinate swimmer to achieve controlled propulsion.

The contents of this paper are as follows.
In the next section, we formulate the stochastic thermodynamics of scallop motion, which contains important notions characterizing scallop motion driven by the BSN curvature.
In Sec. \ref{sec:explicit_calculations}, we present explicit formulas for the harmonic potential.
In Sec. \ref{sec:numerical_demonstrations}, we numerically demonstrate the validity of our theoretical predictions.
In Sec. \ref{sec:discussion}, we discuss our results.
In Sec. \ref{sec:conclusion}, we conclude our results with some remarks.
In the Appendices, we present some details to support the main text.

\section{Stochastic Thermodynamics of Scallop Motion}
\label{Sec:Stochastic}

\subsection{Graphical Summary of the Theoretical Framework}
\label{subsec:graphical_summary}

\begin{figure}
    \centering
    \includegraphics[width=0.8\linewidth]{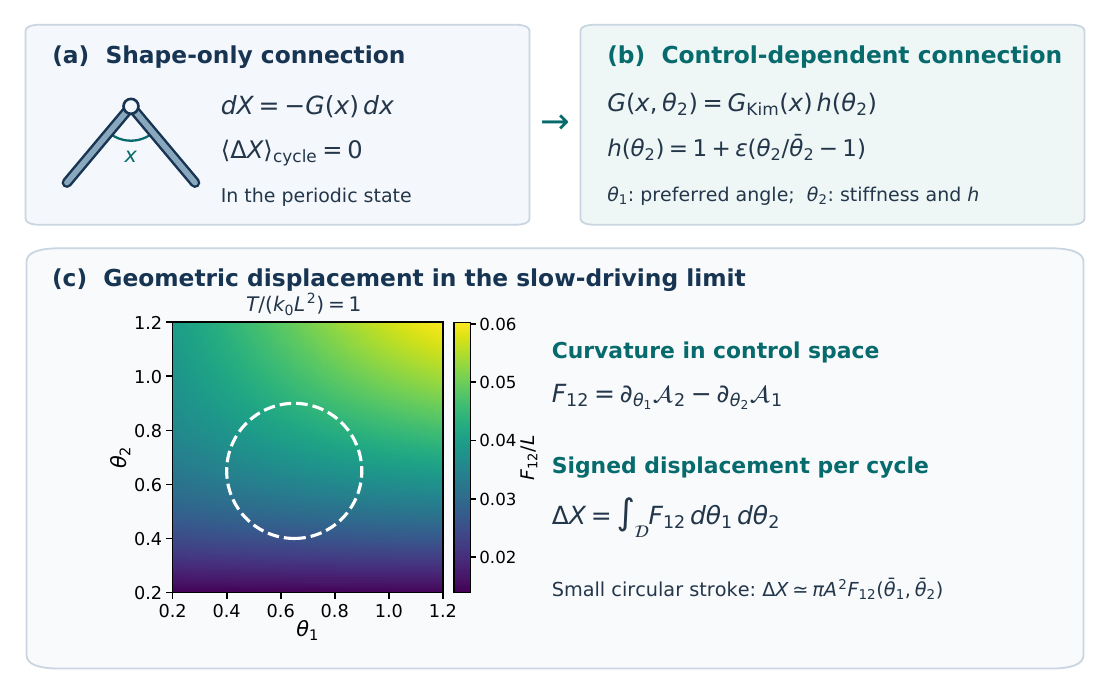}
    \caption{
Schematic illustration of geometric propulsion induced by control-dependent hydrodynamics.
For a conventional one-degree-of-freedom scallop, the hydrodynamic connection depends only on the instantaneous opening angle \(x\), \(G=G(x)\). 
The translational displacement is then an exact differential, \(dX=-G(x)dx\), and hence vanishes over any closed reciprocal cycle, \(\Delta X=0\), in accordance with Purcell's scallop theorem. We introduce a minimal phenomenological extension in which the hydrodynamic connection also depends on the second control parameter,
\(G(x,\theta_2)=G_{\rm Kim}(x)[1+\epsilon(\theta_2/\overline{\theta_2}-1)]\),
where \(G_{\rm Kim}(x)\) represents the angle-dependent hydrodynamic coupling~\cite{Kim1986} and \(\epsilon\) measures the strength of the additional control dependence. 
Together with the stochastic shape dynamics driven by the two control parameters \((\theta_1,\theta_2)\), this produces a nonzero geometric curvature \(F_{12}\) in control space. 
A cyclic protocol enclosing a finite curvature flux therefore generates a finite net displacement,
\(\Delta X=\iint_{\Sigma}F_{12} d\theta_1d\theta_2\neq0\).
For sufficiently small cycles, the displacement is proportional to the enclosed area, while the limit \(\epsilon\to0\) continuously recovers the conventional scallop theorem. Thermal fluctuations modify the stochastic shape distribution and suppress the propulsion at high temperature, but are not the origin of propulsion in the present phenomenological model.
The curvature map in panel (c) uses the same numerical data and circular protocol as Fig. \ref{fig:scallop-curv-ref}.
} 
    \label{fig:schematic}
\end{figure}

Figure~\ref{fig:schematic} summarizes the central physical mechanism
considered in this study. For a conventional one-degree-of-freedom
scallop, the hydrodynamic connection depends only on the instantaneous
shape coordinate. By Eq.~\eqref{eq(1.1)}, the displacement associated
with shape motion is an exact differential. It vanishes for a closed
shape trajectory and, for stochastic shape motion, its mean vanishes
over a control cycle in the periodic state.

The essential extension introduced here is the explicit dependence of
the hydrodynamic response on an external control parameter. The
corresponding force-free kinematics is given by Eq.~\eqref{eq:controlled_kinematics},
and the minimal phenomenological realization adopted in this paper is
Eq.~\eqref{G(x,theta_2)}. In the limit $\epsilon\to0$, this
control dependence disappears and the conventional scallop theorem is
recovered.

The stochastic shape coordinate evolves under two external controls
$\bm{\theta}=(\theta_1,\theta_2)^T$. A slow closed modulation of these
controls generates a geometric connection $\mathcal A_i(\bm{\theta})$
in control space. We define its curvature by
\begin{equation}
F_{12}(\bm{\theta})
=
\partial_{\theta_1}\mathcal A_2
-
\partial_{\theta_2}\mathcal A_1.
\label{eq:curvature_definition}
\end{equation}
The net quasistatic displacement over a closed control protocol
$\mathcal C$ is
\begin{equation}
\Delta X
=
\oint_{\mathcal C}
\mathcal A_i(\bm{\theta})\,d\theta_i
=
\iint_{\mathcal D}
F_{12}(\bm{\theta})\,d\theta_1d\theta_2,
\label{eq:net_displacement_geometric}
\end{equation}
where $\mathcal D$ is an oriented surface bounded by $\mathcal C$.

The second geometric structure considered in this paper is the
thermodynamic metric $g_{ij}(\bm{\theta})$, which characterizes the
leading finite-time entropy production. Thus, the antisymmetric
curvature $F_{12}$ controls geometric propulsion, whereas the
symmetric metric $g_{ij}$ controls the thermodynamic cost of driving.
The following subsections derive these two structures without
identifying thermal fluctuations themselves as the origin of
propulsion.

\subsection{Hydrodynamic Reduction}
\label{subsec:hydrodynamic_reduction}

We first summarize the hydrodynamic reduction required for both the shape dynamics and the translational motion. 
The two mechanical coordinates are the axial hinge coordinate $X$ and the opening angle $x\in\Omega$. Their velocities are denoted by $U=\dot X$ and $\dot x$, respectively.

We restrict attention to axial motion with a fixed or constrained orientation in a setting invariant under translations along the axis. 
The displacement reported here is signed along that axis; it is not a directed laboratory-frame mean for an isotropically oriented ensemble of freely rotating swimmers.

At the level of these generalized coordinates, the resistance
relation is
\begin{equation}
\begin{pmatrix}
F_{\mathrm{prop}}\\[2pt]
\tau_x
\end{pmatrix}
=
\begin{pmatrix}
\zeta_{XX}(x,\bm{\theta})
&
\zeta_{Xx}(x,\bm{\theta})
\\
\zeta_{xX}(x,\bm{\theta})
&
\zeta_{xx}(x,\bm{\theta})
\end{pmatrix}
\begin{pmatrix}
U\\[2pt]
\dot x
\end{pmatrix},
\label{eq:resistance_matrix}
\end{equation}
where $F_{\mathrm{prop}}$ and $\tau_x$ are the generalized forces
conjugate to $X$ and $x$, respectively. Lorentz reciprocity gives
\begin{equation}
\zeta_{Xx}(x,\bm{\theta})
=
\zeta_{xX}(x,\bm{\theta}).
\label{eq:reciprocity}
\end{equation}

For a force-free swimmer, $F_{\mathrm{prop}}=0$, and
Eq.~\eqref{eq:resistance_matrix} reduces to the kinematic relation
already introduced in Eq.~\eqref{eq:controlled_kinematics}, with
\begin{equation}
G(x,\bm{\theta})
=
\frac{
\zeta_{Xx}(x,\bm{\theta})
}{
\zeta_{XX}(x,\bm{\theta})
}.
\label{eq:G_resistance}
\end{equation}
Thus, Eq.~\eqref{eq:G_resistance} provides the hydrodynamic meaning
of the connection appearing in Sec.~I. The specific weak control
dependence employed below is the phenomenological model of Eq.~\eqref{G(x,theta_2)}.

Eliminating the translational velocity from the second row of
Eq.~\eqref{eq:resistance_matrix} gives the effective resistance for
the internal shape motion,
\begin{equation}
\zeta_{\mathrm{eff}}(x,\bm{\theta})
=
\zeta_{xx}(x,\bm{\theta})
-
\frac{\zeta_{Xx}^2(x,\bm{\theta})}
{\zeta_{XX}(x,\bm{\theta})},
\label{eq:zeta_eff}
\end{equation}
and hence the free-swimming effective mobility
\begin{equation}
M_{\mathrm{eff}}(x,\bm{\theta})
=
\frac{1}{\zeta_{\mathrm{eff}}(x,\bm{\theta})}.
\label{eq:M_eff}
\end{equation}
For comparison, the clamped mobility is
\begin{equation}
M_{\mathrm{cl}}(x,\bm{\theta})
=
\frac{1}{\zeta_{xx}(x,\bm{\theta})}, 
\label{eq:M_cl}
\end{equation}
Positive definiteness of the resistance matrix implies
\begin{equation}
M_{\mathrm{eff}}(x,\bm{\theta})
\geq
M_{\mathrm{cl}}(x,\bm{\theta}).
\label{eq:mobility_inequality}
\end{equation}

The positive-definite resistance description applies, for example,
to a genuinely three-dimensional or confined system. The comparison
between force-free and clamped motion refers to the same resistance
matrix, including any dependence on the controls. We use the
unbounded two-dimensional result of Kim et al.~\cite{Kim1986} only for
the force-free connection, not for the resistance to externally
forced translation.

The two hydrodynamic quantities $M_{\mathrm{eff}}$ and $G$ therefore
have distinct roles. The former determines the stochastic relaxation
of the internal coordinate, whereas the latter converts shape motion
into translation.

For the shape dependence of the translational connection, we use the
finite-hinged-plate calculation of Kim et al.~\cite{Kim1986}. If
$\alpha=x/2$ denotes the half-opening angle and $\Omega$ is their
angular velocity, their force-free result defines
\begin{equation}
\frac{G_{\mathrm{Kim}}(x)}{L}
=
\frac{1}{2}
\left.\frac{U_{\infty}}{\Omega L}\right|_{\alpha=x/2}.
\label{eq:G_Kim}
\end{equation}
Here $U_{\infty}$ is the far-fluid velocity in their hinge-fixed frame;
our positive $X$ direction is opposite to their positive axial direction.
The numerical implementation uses an interpolation of the data in
Ref.~\cite{Kim1986}. Details of the hydrodynamic construction are
given in Appendix~\ref{app:hydrodynamic_reduction}.

\subsection{Stochastic Shape Dynamics and Geometric Response}
\label{II_C}

We now formulate the stochastic dynamics of the internal coordinate
and introduce all dimensionless quantities used in the slow-driving
analysis.

The internal potential is taken to be
\begin{equation}
H(x;\bm{\theta})
=
\frac{1}{2}
k_0L^2\theta_2
(x-\theta_1)^2,
\label{eq:two_param_harmonic}
\end{equation}
where $\theta_1$ determines the preferred opening angle and
$\theta_2$ determines the effective stiffness.

We consider periodic protocols
\begin{align}
\theta_1(t)
&=
\bar{\theta}_1
+
A_1\sin(\omega_1t),
\nonumber\\
\theta_2(t)
&=
\bar{\theta}_2
+
A_2\sin(\omega_2t+\phi).
\label{eq:trajectory}
\end{align}
A closed protocol of period $\tau_p=2\pi/\omega_0$ is obtained when
\begin{equation}
n_i
:=
\frac{\omega_i}{\omega_0}
\in\mathbb Z,
\qquad
i=1,2.
\label{eq:frequency_ratio}
\end{equation}

On the fixed shape interval $\Omega=[x_-,x_+]$ we impose reflecting
boundaries, $J_x(x_-,t)=J_x(x_+,t)=0$. The physical Smoluchowski equation is
\begin{equation}
\frac{\partial\rho(x,t)}{\partial t}
=
\mathcal L_{\bm{\theta}(t)}
\rho(x,t),
\label{eq:smoluchowski_physical}
\end{equation}
with
\begin{equation}
\mathcal L_{\bm{\theta}}\rho
=
-\partial_xJ_x[\rho],
\label{eq:L_current}
\end{equation}
and
\begin{equation}
J_x[\rho]
=
-M_{\mathrm{eff}}(x,\bm{\theta})
\left[
\partial_xH(x;\bm{\theta})\,\rho
+
T\,\partial_x\rho
\right].
\label{eq:physical_current}
\end{equation}

For fixed $\bm{\theta}$, detailed balance gives
\begin{equation}
J_x[\rho^{\mathrm{ss}}]=0,
\end{equation}
and hence
\begin{equation}
\rho^{\mathrm{ss}}(x;\bm{\theta})
=
\frac{1}{Z(\bm{\theta})}
\exp\left[
-\frac{H(x;\bm{\theta})}{T}
\right],
\qquad
Z(\bm{\theta})
=
\int_{\Omega}dx\,
e^{-H(x;\bm{\theta})/T}.
\label{eq:rho_ss_exact}
\end{equation}

We now introduce the dimensionless slow-driving quantities. The
dimensionless operation speed and protocol time are defined by
\begin{equation}
\delta
:=
\frac{1}{\Gamma\tau_p}
\ll1,
\qquad
s
:=
\frac{t}{\tau_p},
\qquad
0\leq s\leq1,
\label{eq:delta_s}
\end{equation}
where $\Gamma$ is a characteristic relaxation rate.

We further define the dimensionless Smoluchowski operator and mobility
by
\begin{equation}
\widetilde{\mathcal L}_{\bm{\theta}}
:=
\frac{\mathcal L_{\bm{\theta}}}{\Gamma},
\qquad
\widetilde M(x,\bm{\theta})
:=
\frac{T M_{\mathrm{eff}}(x,\bm{\theta})}{\Gamma}.
\label{eq:dimensionless_L_M}
\end{equation}
The dimensionless evolution equation is therefore
\begin{equation}
\delta
\frac{\partial\rho}{\partial s}
=
\widetilde{\mathcal L}_{\bm{\theta}}\rho,
\label{eq:dimensionless_smoluchowski}
\end{equation}
where
\begin{equation}
\widetilde{\mathcal L}_{\bm{\theta}}\rho
=
\partial_x
\left\{
\widetilde M(x,\bm{\theta})
\left[
\frac{\partial_xH(x;\bm{\theta})}{T}\rho
+
\partial_x\rho
\right]
\right\}.
\label{eq:dimensionless_generator}
\end{equation}

We work in the periodic state after transients have decayed and
expand the probability density as
\begin{equation}
\rho(x,s)
=
\rho^{\mathrm{ss}}(x;\bm{\theta}(s))
+
\delta\rho^{(1)}(x,s)
+
O(\delta^2).
\label{eq:rho_slow_expansion}
\end{equation}
At first order,
\begin{equation}
\widetilde{\mathcal L}_{\bm{\theta}}
\rho^{(1)}
=
\frac{d\theta_i}{ds}
\frac{\partial\rho^{\mathrm{ss}}}{\partial\theta_i}.
\label{eq:first_order_density}
\end{equation}

We define the density-response functions
\begin{equation}
\chi_i(x;\bm{\theta})
:=
\widetilde{\mathcal L}_{\bm{\theta}}^{+}
\left(
\frac{\partial\rho^{\mathrm{ss}}}
{\partial\theta_i}
\right),
\label{eq:chi_def}
\end{equation}
where
$\widetilde{\mathcal L}_{\bm{\theta}}^{+}$ is the generalized inverse
on the zero-integral subspace. Then
\begin{equation}
\rho^{(1)}(x,s)
=
\frac{d\theta_i}{ds}
\chi_i(x;\bm{\theta}).
\label{eq:rho1_chi}
\end{equation}

For the one-dimensional problem, it is useful to introduce the
cumulative stationary distribution
\begin{equation}
\mathcal R(x;\bm{\theta})
:=
\int_{x_-}^{x}
dy\,
\rho^{\mathrm{ss}}(y;\bm{\theta})
\label{eq:cumulative_distribution}
\end{equation}
and its parameter derivatives
\begin{equation}
S_i(x;\bm{\theta})
:=
\frac{\partial\mathcal R(x;\bm{\theta})}
{\partial\theta_i}
=
\int_{x_-}^{x}
dy\,
\frac{\partial\rho^{\mathrm{ss}}(y;\bm{\theta})}
{\partial\theta_i}.
\label{eq:S_i_def}
\end{equation}

The dimensionless probability current,
\begin{equation}
\widetilde J_x
:=
\frac{J_x}{\Gamma},
\label{eq:J_tilde_def}
\end{equation}
has the expansion
\begin{equation}
\widetilde J_x
=
-\delta
\frac{d\theta_i}{ds}
S_i(x;\bm{\theta})
+
O(\delta^2).
\label{eq:J_tilde_first_order}
\end{equation}
This follows directly by integrating the dimensionless continuity
equation from the reflecting boundary $x_-$ to $x$.

The same result follows from the density response:
\begin{equation}
-\widetilde M(x,\bm{\theta})
\left[
\frac{\partial_xH}{T}
\chi_i
+
\partial_x\chi_i
\right]
=
-S_i.
\label{eq:chi_S_relation}
\end{equation}
Thus, $\chi_i$ describes the first-order correction to the density,
whereas $S_i$ is the corresponding integrated current response. The
two quantities have distinct meanings and will not be denoted by the
same symbol below.

We next combine the stochastic current with the hydrodynamic
kinematics of Eq.~\eqref{eq:controlled_kinematics}.
The mean
translational velocity is
\begin{equation}
\langle U(t)\rangle
=
-\int_{\Omega}dx\,
G(x,\bm{\theta})J_x(x,t).
\label{eq:mean_velocity}
\end{equation}
Using Eq.~\eqref{eq:J_tilde_first_order} and
$\dot\theta_i=\Gamma\delta\,d\theta_i/ds$, we obtain
\begin{equation}
\langle U(t)\rangle
=
\dot\theta_i
\mathcal A_i(\bm{\theta})
+
O(\delta^2),
\label{eq:mean_velocity_geometric}
\end{equation}
where the geometric connection appearing in
Eq.~\eqref{eq:net_displacement_geometric} is
\begin{equation}
\mathcal A_i(\bm{\theta})
=
\int_{\Omega}dx\,
G(x,\bm{\theta})
S_i(x;\bm{\theta}).
\label{eq:A_i_explicit}
\end{equation}

Using Eq.~\eqref{eq:curvature_definition}, the curvature becomes
\begin{align}
F_{12}(\bm{\theta})
=
\int_{\Omega}dx\,
\bigg[
&
\frac{\partial G(x,\bm{\theta})}{\partial\theta_1}
S_2(x;\bm{\theta})-
\frac{\partial G(x,\bm{\theta})}{\partial\theta_2}
S_1(x;\bm{\theta})
\bigg].
\label{eq:F12_general}
\end{align}
Terms containing
$\partial_{\theta_1}\partial_{\theta_2}\mathcal R$
cancel in the antisymmetric combination.

Equation~\eqref{eq:F12_general} makes the stochastic version of the
scallop constraint explicit. If the hydrodynamic connection has no
explicit control dependence, $G=G(x)$, then
\begin{equation}
F_{12}(\bm{\theta})=0.
\label{eq:F_zero_Gx}
\end{equation}
The absence of mean propulsion for $G=G(x)$ also holds at any
driving rate in the periodic state. If $Q'(x)=G(x)$, the continuity
equation and the reflecting boundaries give
$\langle U\rangle=-d\langle Q\rangle/dt$; its period integral is zero
when $\rho(x,t+\tau_p)=\rho(x,t)$. A transient can change
$\langle Q\rangle$, and zero mean propulsion does not exclude axial
Brownian diffusion if translational noise is included.

For the factorized connection in Eq.~\eqref{G(x,theta_2)}, write
$G(x,\theta_2(t))=h(t)Q'(x)$, where $Q'=G_{\rm Kim}$ and
$h(t)=1+\epsilon[\theta_2(t)/\bar{\theta}_2-1]$. Define
\[
m_Q(t)=\int_\Omega Q(x)\rho(x,t)\,dx.
\]
The continuity equation and the zero boundary currents give
\[
\dot m_Q=\int_\Omega Q'(x)J_x(x,t)\,dx,\qquad
\langle U(t)\rangle=-h(t)\dot m_Q(t),
\]
where the second equality follows from Eq.~\eqref{eq:mean_velocity}.
For a drive of period $\tau_p$, integration by parts in the
periodic state gives the exact mean displacement
\begin{equation}
\Delta X=\int_0^{\tau_p}m_Q(t)\dot h(t)\,dt
        =\oint m_Q\,dh .
\label{eq:finite-rate-displacement}
\end{equation}
The identity is exact at any driving rate within the instantaneous Stokes description; it does not include fluid inertia or hydrodynamic memory.
The boundary term vanishes because both $h$ and $m_Q$ are periodic.
This identity requires neither slow driving relative to shape
relaxation nor an instantaneous equilibrium shape distribution.
It does assume the kinematics of Eq.~\eqref{eq:controlled_kinematics},
with no additional flow generated directly by varying the boundary
control. For a calibrated $Q$ and $h$, the right-hand side can be
obtained from time-resolved angle statistics without fitting a
propulsion law separately at each driving frequency.

\subsection{Excess Entropy Production and the Metric}
\label{II_D}

We now turn to the thermodynamic cost of slow driving. Because the
instantaneous stationary state satisfies detailed balance, the
quasistatic entropy production vanishes.

For completeness, the equilibrium free energy is
\begin{equation}
\mathcal F(\bm{\theta})
=
-T\ln Z(\bm{\theta}).
\label{eq:free_energy}
\end{equation}
The quasistatic work satisfies
\begin{equation}
dW_{\mathrm{qs}}
=
d\mathcal F,
\end{equation}
and therefore
\begin{equation}
W^{(0)}=0,
\qquad
Q^{(0)}=0,
\qquad
\Sigma^{(0)}=0
\label{eq:zeroth_thermodynamics}
\end{equation}
for a closed isothermal cycle.

The entropy-production rate of the reduced shape dynamics (excluding
any cost of independently operating the hydrodynamic actuator) is
\begin{equation}
\dot\Sigma(t)
=
\int_{\Omega}dx\,
\frac{
J_x(x,t)^2
}{
T M_{\mathrm{eff}}(x,\bm{\theta})\rho(x,t)
}
\geq0.
\label{eq:entropy_production_rate}
\end{equation}
Using
$J_x=\Gamma\widetilde J_x$,
$T M_{\mathrm{eff}}=\Gamma\widetilde M$,
and Eq.~\eqref{eq:J_tilde_first_order}, we find
\begin{equation}
\dot\Sigma
=
\Gamma\delta^2
\frac{d\theta_i}{ds}
\frac{d\theta_j}{ds}
\int_{\Omega}dx\,
\frac{
S_i(x;\bm{\theta})S_j(x;\bm{\theta})
}{
\widetilde M(x,\bm{\theta})
\rho^{\mathrm{ss}}(x;\bm{\theta})
}
+
O(\delta^3).
\label{eq:Sigma_dot_expansion}
\end{equation}

We therefore define the dimensionless thermodynamic metric once and
for all by
\begin{equation}
g_{ij}(\bm{\theta})
:=
\int_{\Omega}dx\,
\frac{
S_i(x;\bm{\theta})S_j(x;\bm{\theta})
}{
\widetilde M(x,\bm{\theta})
\rho^{\mathrm{ss}}(x;\bm{\theta})
}.
\label{eq:metric_tensor}
\end{equation}
This representation makes it explicit that $g_{ij}=g_{ji}$ and that
the metric is positive semidefinite.

Because $dt=\tau_p ds=(\Gamma\delta)^{-1}ds$, the total entropy
production over one period is
\begin{equation}
\Sigma
=
\delta\Sigma^{(1)}
+
O(\delta^2),
\label{eq:Sigma_expansion}
\end{equation}
where the coefficient of the leading entropy production is defined by
\begin{equation}
\Sigma^{(1)}
:=
\int_0^1ds\,
g_{ij}(\bm{\theta}(s))
\frac{d\theta_i}{ds}
\frac{d\theta_j}{ds}
\geq0.
\label{eq:Sigma_first_order}
\end{equation}
The leading excess entropy production is therefore
\begin{equation}
\Sigma_{\mathrm{ex}}
=
\delta\Sigma^{(1)}.
\label{eq:Sigma_ex}
\end{equation}

There is no independent reversible contribution to
$\Sigma^{(1)}$. Reversible changes of the system entropy and entropy
flow to the bath can occur during the quasistatic protocol, but they
cancel in the total entropy production. The first nonvanishing total
entropy production is therefore the irreversible contribution in
Eq.~\eqref{eq:Sigma_expansion}.

For the present single-bath isothermal cycle,
\begin{equation}
W_{\mathrm{ex}}
=
T\Sigma_{\mathrm{ex}},
\qquad
Q_{\mathrm{ex}}
=
-T\Sigma_{\mathrm{ex}},
\label{eq:WQ_excess}
\end{equation}
where heat absorbed by the system is taken to be positive.

\subsection{Thermodynamic Length and Spectral Relaxation}
\label{II_E}

The metric in Eq.~\eqref{eq:metric_tensor} defines the thermodynamic
line element
\begin{equation}
d\ell_{\mathrm{th}}^2
=
g_{ij}(\bm{\theta})
\,d\theta_i d\theta_j.
\label{eq:thermodynamic_line_element}
\end{equation}
The thermodynamic length of a closed path $\mathcal C$ is defined,
once and for all, by
\begin{equation}
\mathcal L_{\mathrm{th}}
:=
\oint_{\mathcal C}
\sqrt{
g_{ij}(\bm{\theta})\,d\theta_i d\theta_j
}.
\label{eq:thermodynamic_length}
\end{equation}
Equivalently,
\begin{equation}
\mathcal L_{\mathrm{th}}
=
\int_0^1ds\,
v_{\mathrm{th}}(s),
\qquad
v_{\mathrm{th}}(s)
=
\sqrt{
g_{ij}(\bm{\theta}(s))
\frac{d\theta_i}{ds}
\frac{d\theta_j}{ds}
}.
\label{eq:thermodynamic_speed}
\end{equation}

The Cauchy--Schwarz inequality gives
\begin{equation}
\Sigma^{(1)}
\geq
\mathcal L_{\mathrm{th}}^2.
\label{eq:CS_dimensionless}
\end{equation}
Equality holds if and only if
\begin{equation}
v_{\mathrm{th}}(s)
=
\mathcal L_{\mathrm{th}}
\end{equation}
is constant along the protocol. Thus, for a prescribed path, the
leading dissipation is minimized by uniform motion in thermodynamic
arclength.

The deviation from the optimal parametrization is
\begin{equation}
\Sigma^{(1)}
-
\mathcal L_{\mathrm{th}}^2
=
\int_0^1ds\,
\left[
v_{\mathrm{th}}(s)
-
\mathcal L_{\mathrm{th}}
\right]^2
\geq0.
\label{eq:thermodynamic_speed_variance}
\end{equation}

Finally, we relate the slow-driving response to the relaxation
spectrum. Let the nonstationary right and left eigenfunctions satisfy
\begin{align}
\widetilde{\mathcal L}_{\bm{\theta}}r_n
&=
-\widetilde\lambda_n(\bm{\theta})r_n,
\nonumber\\
\widetilde{\mathcal L}_{\bm{\theta}}^\dagger \ell_n
&=
-\widetilde\lambda_n(\bm{\theta})\ell_n,
\qquad
n\geq1,
\label{eq:spectral_modes}
\end{align}
with
$\langle \ell_m|r_n\rangle=\delta_{mn}$.
The generalized inverse appearing in
Eq.~\eqref{eq:chi_def} is
\begin{equation}
\widetilde{\mathcal L}_{\bm{\theta}}^{+}
=
-
\sum_{n\geq1}
\frac{
|r_n\rangle\langle \ell_n|
}{
\widetilde\lambda_n(\bm{\theta})
}.
\label{eq:pseudoinverse_spectral}
\end{equation}

A small nonzero relaxation eigenvalue can therefore enhance the
density response $\chi_i$. However, the thermodynamic metric is not
determined by the spectral gap alone: the contribution of each mode
also depends on its overlap with
$\partial_{\theta_i}\rho^{\mathrm{ss}}$.
A region of small $\widetilde M$ can increase the cost of any
protocol that carries appreciable shape probability current through
that region, as is explicit in Eq.~\eqref{eq:metric_tensor}.

The two geometric structures introduced in this section are therefore
distinct. The curvature $F_{12}$ describes the displacement generated
by a closed control cycle in the slow-driving limit, whereas the metric $g_{ij}$ and the
associated thermodynamic length describe its dissipative cost.

\section{Explicit Results for the Harmonic Model}
\label{sec:explicit_calculations}

We now specialize the general formulas in the previous section to the harmonic potential in Eq.~\eqref{eq:two_param_harmonic}. 
No new response functions or thermodynamic metrics are introduced in this section.

\subsection{Geometric Curvature and Area Law}
\label{subsec:harmonic_geometric_curvature}

We impose reflecting boundaries at the fixed endpoints \(x_-\) and
\(x_+\) of \(\Omega\). Let \(a=k_0L^2/(2T)\), \(q=\sqrt{a\theta_2}\),
and \(z_j=q(x_j-\theta_1)\) for \(j=x,-,+\), with \(x_x=x\).
The normalized harmonic equilibrium and its cumulative distribution are
\begin{align}
 \rho^{\mathrm{ss}}(x;\boldsymbol\theta)
 &=\frac{2q e^{-z_x^2}}
 {\sqrt{\pi}[\operatorname{erf}(z_+)-\operatorname{erf}(z_-)]},
 \label{eq:scallop-finite-rho}\\
 \mathcal R(x;\boldsymbol\theta)
 &=\frac{\operatorname{erf}(z_x)-\operatorname{erf}(z_-)}
 {\operatorname{erf}(z_+)-\operatorname{erf}(z_-)} ,
\end{align}
where \(\operatorname{erf}(z):=\frac{2}{\sqrt{\pi}}\int_0^z dt e^{-t^2}\) is the error function.
Writing \(\rho_\pm=\rho^{\mathrm{ss}}(x_\pm)\), we obtain
\begin{align}
 S_1(x)&=-\rho^{\mathrm{ss}}(x)+(1-\mathcal R)\rho_-+\mathcal R\rho_+,
 \label{eq:scallop-finite-S1}\\
 S_2(x)&=\frac{(x-\theta_1)\rho^{\mathrm{ss}}(x)
       -(1-\mathcal R)(x_--\theta_1)\rho_- -\mathcal R(x_+-\theta_1)\rho_+}{2\theta_2}.
 \label{eq:scallop-finite-S2}
\end{align}
Both functions vanish at both endpoints, consistent with the reflecting
boundary conditions. The full-line Gaussian expressions follow when
the stationary distribution is well localized in the interior; we do
not use that approximation in the figures.

With \(G(x,\theta_2)=G_{\rm Kim}(x)
[1+\epsilon(\theta_2/\bar\theta_2-1)]\), the curvature is
\begin{equation}
 F_{12}=-\frac{\epsilon}{\bar\theta_2}
 \int_{x_-}^{x_+}G_{\rm Kim}(x)S_1(x)\,dx .
 \label{eq:scallop-finite-curv}
\end{equation}
Raising \(\theta_1\) shifts weight to larger opening angles. If \(Y\)
denotes the random opening angle, then
\[
 S_1(x)=2a\theta_2\operatorname{Cov}_{\rm ss}[\mathbf 1_{\{Y\le x\}},Y]\le0, 
\]
where $\mathbf{1}_{\{Y\le x\}}$ denotes the indicator, equal to one when $Y\le x$ and zero otherwise.
Here $\operatorname{Cov}_{\rm ss}$ denotes covariance at fixed $\boldsymbol{\theta}$ with respect to the stationary distribution:
\begin{equation}
\operatorname{Cov}_{\rm ss}\!\bigl(f(Y),g(Y)\bigr)
:=\langle f(Y)g(Y)\rangle_{\rm ss}
 -\langle f(Y)\rangle_{\rm ss}\langle g(Y)\rangle_{\rm ss},
\end{equation}
where
$\langle h(Y)\rangle_{\rm ss}
:=\int_{x_-}^{x_+} h(y)\rho^{\mathrm{ss}}(y;\boldsymbol{\theta})\,dy$.
Thus a positive \(G_{\rm Kim}\) gives a nonnegative curvature for
\(\epsilon>0\) in the stated coordinate convention.
If \(\theta_1\) is in the interior, then
\(F_{12}\to(\epsilon/\bar\theta_2)G_{\rm Kim}(\theta_1)\) as \(T\to0\).
On a fixed finite interval at high temperature the equilibrium approaches
the uniform distribution and \(S_1=O(T^{-1})\), so
\(F_{12}=O(T^{-1})\). The full-line approximation does not predict this
finite-interval high-temperature behavior.

For a sufficiently small circular protocol of radius $A$ centered at
$\bar{\bm{\theta}}=(\bar{\theta}_1,\bar{\theta}_2)$,
the curvature is approximately constant over the enclosed region.
Equation~\eqref{eq:net_displacement_geometric} then gives
\begin{equation}
\Delta X
\simeq
\pi A^2
F_{12}(\bar{\bm{\theta}}).
\label{eq:area_law}
\end{equation}
The displacement therefore obeys the geometric area law
$\Delta X\propto A^2$ in the small-loop regime.

Moreover,
$F_{12}\propto\epsilon$, so that
\begin{equation}
\lim_{\epsilon\to0}
\Delta X
=
0.
\label{eq:epsilon_zero_displacement}
\end{equation}
The conventional scallop theorem is continuously recovered when the
control dependence of the hydrodynamic connection is removed.

\subsection{Low-Temperature Limit and Deterministic Geometric Propulsion}
\label{subsec:zero_temperature_limit}

The low-temperature limit provides a direct interpretation of the
propulsion mechanism. As $T$ decreases, the stationary distribution
in Eq.~\eqref{eq:scallop-finite-rho} becomes localized around
$x=\theta_1$, provided $\theta_1$ lies in the interior of $\Omega$.
Formally,
\begin{equation}
\rho^{\mathrm{ss}}(x;\bm{\theta})
\xrightarrow[T\to0]{}
\delta(x-\theta_1).
\label{eq:rho_zero_temperature}
\end{equation}
In this limit the boundary terms in $S_1$ vanish, so
Eq.~\eqref{eq:scallop-finite-curv} yields
\begin{equation}
F_{12}(\bm{\theta})
\xrightarrow[T\to0]{}
\frac{\epsilon}{\bar{\theta}_2}
G_{\mathrm{Kim}}(\theta_1).
\label{eq:F_zero_temperature}
\end{equation}
The curvature thus remains finite in the deterministic
vanishing-noise limit.

This result is consistent with Purcell's theorem. In the conventional
case described by Eq.~\eqref{eq(1.1)}, the hydrodynamic connection has no
independent control dependence and the displacement is an exact
differential. In the present model, however, Eq.~\eqref{G(x,theta_2)}
contains the independent control $\theta_2$. When
$x\to\theta_1$ in the deterministic limit, Eq.~\eqref{eq:controlled_kinematics}
implies
\begin{equation}
dX
=
-G(\theta_1,\theta_2)\,d\theta_1,
\label{eq:deterministic_controlled_motion}
\end{equation}
which is not generally an exact differential in the
two-dimensional control space.

Thus, thermal noise is neither necessary for propulsion nor a
mechanism that circumvents the scallop theorem. The essential
ingredient is the control dependence of the hydrodynamic response.
Finite temperature changes the current response $S_1$ and thus
the weighting of $G_{\mathrm{Kim}}$ in
Eq.~\eqref{eq:scallop-finite-curv}.

\subsection{Clamped-Mobility Upper Bound}
\label{subsec:clamped_mobility_bound}

The exact thermodynamic metric follows from
Eq.~\eqref{eq:metric_tensor} with the free-swimming mobility
$\widetilde M(x,\bm{\theta})$. In practice, however, the complete resistance
matrix required to evaluate $M_{\mathrm{eff}}$ may not be available.
The exact clamped mobility $M_{\mathrm{cl}}$ in
Eq.~\eqref{eq:M_cl} is defined from the same resistance matrix and
can be used for a conditional comparison.

We define its dimensionless counterpart by
\begin{equation}
\widetilde M_{\mathrm{cl}}(x,\bm{\theta})
=
\frac{T M_{\mathrm{cl}}(x,\bm{\theta})}{\Gamma}.
\label{eq:Mcl_tilde}
\end{equation}
The corresponding clamped metric is
\begin{equation}
g^{\mathrm{cl}}_{ij}(\bm{\theta})
=
\int_{\Omega}dx\,
\frac{
S_i(x;\bm{\theta})S_j(x;\bm{\theta})
}{
\widetilde M_{\mathrm{cl}}(x,\bm{\theta})
\rho^{\mathrm{ss}}(x;\bm{\theta})
}.
\label{eq:clamped_metric}
\end{equation}

Because
$M_{\mathrm{eff}}\geq M_{\mathrm{cl}}$,
the difference between the clamped and free-swimming metrics is
positive semidefinite. More precisely, for any vector
$\bm v=(v_1,v_2)^T$,
\begin{equation}
v_i g_{ij}v_j
\leq
v_i g^{\mathrm{cl}}_{ij}v_j.
\label{eq:metric_Loewner_bound}
\end{equation}
This is the appropriate tensorial statement of the upper bound; it
does not require a componentwise inequality for the off-diagonal
elements.

Consequently, for any prescribed protocol,
\begin{equation}
\Sigma^{(1)}
\leq
\Sigma_{\mathrm{cl}}^{(1)},
\label{eq:entropy_clamped_bound}
\end{equation}
where
\begin{equation}\label{eq:Sigma_cl_numerical}
\Sigma_{\mathrm{cl}}^{(1)}
=
\int_0^1ds\,
g^{\mathrm{cl}}_{ij}(\bm{\theta}(s))
\frac{d\theta_i}{ds}
\frac{d\theta_j}{ds}.
\end{equation}
This bound requires the exact clamped resistance of the same system
and holds for the reduced shape-sector dissipation. In particular, it
does not imply a numerical bound based on the interior-wedge
reference resistance used below. The curvature does not depend on
the shape mobility at leading adiabatic order, at fixed $H$ and $G$.

\section{Numerical Demonstrations and Protocols}
\label{sec:numerical_demonstrations}

We illustrate the slow-driving results in two examples. First we
compute the geometric displacement using the control-dependent
connection. We then evaluate the thermodynamic-length bound for an
explicit reference metric.

\subsection{Geometric curvature and cycle displacement}
\begin{figure}[t]
 \centering
 \includegraphics[width=0.58\linewidth]{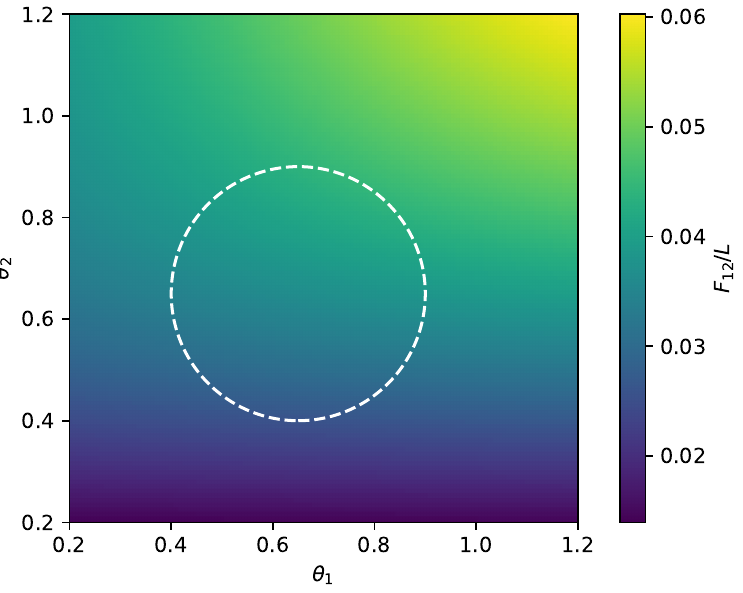}
 \caption{Geometric curvature \(F_{12}/L\) from the normalized
 reflecting harmonic model at \(T/(k_0L^2)=1\). The dashed circle is
 the counterclockwise stroke of radius \(A=0.25\) used in
 Fig.~\ref{fig:scallop-temp-ref}.}
 \label{fig:scallop-curv-ref}
\end{figure}

\begin{figure}[t]
 \centering
 \includegraphics[width=0.58\linewidth]{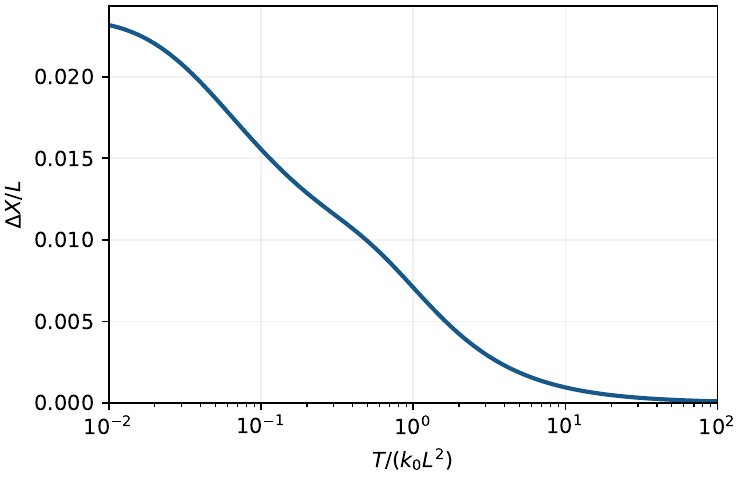}
 \caption{Signed displacement per cycle for the counterclockwise
 stroke with center \((0.65,0.65)\) and radius \(A=0.25\), as a function
 of \(T/(k_0L^2)\).}
 \label{fig:scallop-temp-ref}
\end{figure}

\begin{figure}[t]
 \centering
 \includegraphics[width=0.58\linewidth]{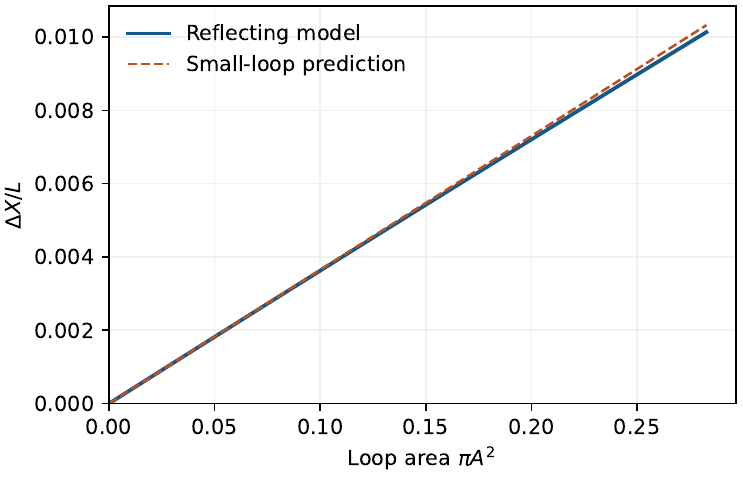}
 \caption{Signed displacement per cycle versus the area of the
 counterclockwise circle at \(T/(k_0L^2)=1\). The dashed line is the
 local area prediction, with no fitted parameter.}
 \label{fig:scallop-area-ref}
\end{figure}

For all numerical calculations we impose reflecting boundaries at
\(x_-=\pi/18\) and \(x_+=\pi\). This interval is a prescribed range of
the model, chosen to coincide with the angular range of the re-digitized
connection. In particular, the upper endpoint uses the exact flat-plate
limit, and no extrapolation to a closed hinge is made.
For Figs.~\ref{fig:scallop-curv-ref}--\ref{fig:scallop-area-ref} we take
\(\epsilon=0.20\) and \((\bar\theta_1,\bar\theta_2)=(0.65,0.65)\).
For comparison, the reference-metric demonstration in the following
subsection uses \(a=200\), or \(T/(k_0L^2)=1/400\).

At leading adiabatic order, replace $m_Q(t)$ in
Eq.~\eqref{eq:finite-rate-displacement} by the stationary expectation
$\langle Q(Y)\rangle_{\rm ss}$. Equivalently, since the current
responses $S_i$ vanish at both boundaries, integration by parts
gives $\mathcal A_i=-h\,\partial_i\langle Q(Y)\rangle_{\rm ss}$.
For the harmonic potential one obtains
\begin{equation}
\Delta X=\frac{\epsilon}{\bar{\theta}_2}
          \oint_{\cal C}\langle Q(Y)\rangle_{\rm ss}\,d\theta_2,
\qquad
F_{12}=\frac{\epsilon}{\bar{\theta}_2}\,
2a\theta_2\operatorname{Cov}_{\rm ss}\!\bigl(Q(Y),Y\bigr).
\label{eq:scallop-numerical-primitive}
\end{equation}
These equivalent line and curvature formulas give independent
numerical checks. Our circular paths are
\((\theta_1,\theta_2)=(\bar\theta_1,\bar\theta_2)+
A(\cos(2\pi s),\sin(2\pi s))\), with counterclockwise orientation.
The plotted displacements are signed, not absolute values.

Figure~\ref{fig:scallop-curv-ref} shows the positive curvature at
\(T/(k_0L^2)=1\). Figure~\ref{fig:scallop-temp-ref} shows that
for the selected stroke the displacement decreases over the displayed
temperature range and remains nonzero as the noise strength is lowered.
On the fixed shape interval it tends to zero at high temperature as
the \(\boldsymbol\theta\)-dependence of the equilibrium density is lost.
Figure~\ref{fig:scallop-area-ref} compares the integrated displacement
with the local prediction \(\pi A^2 F_{12}(\bar{\boldsymbol\theta})\).
The slope is fixed by the curvature rather than fitted to the data.

\subsection{Reference metric and thermodynamic-length bound}

For the numerical demonstration, we use the interior-wedge
resistance per unit span, $\zeta_{\rm ref}^{\rm (per\,span)}(x)$,
given by Eq.~\eqref{eq:scallop-ref-resistance}. We assign the wedge
a fixed out-of-plane reference span $b$ and set
$\zeta_{\rm ref}(x)=b\,\zeta_{\rm ref}^{\rm (per\,span)}(x)$.
The dimensionless reference mobility is
$\widetilde M_{\rm ref}=T/(\Gamma\zeta_{\rm ref})$. 
The metric is
\begin{equation}
 g^{\rm ref}_{ij}=\frac{\Gamma}{T}\int_{x_-}^{x_+}
 \zeta_{\rm ref}(x)\frac{S_i(x)S_j(x)}{\rho^{\mathrm{ss}}(x)}\,dx.
 \label{eq:scallop-reference-metric}
\end{equation}
The exact clamped metric bounds the force-free metric when both
mobilities follow from the same positive-definite resistance matrix.
No such upper bound is asserted for the reference metric.
The reference metric quantifies shape-sector dissipation, not the
cost of operating a separate hydrodynamic actuator.

\begin{figure}[t]
 \centering
 \includegraphics[width=.96\linewidth]{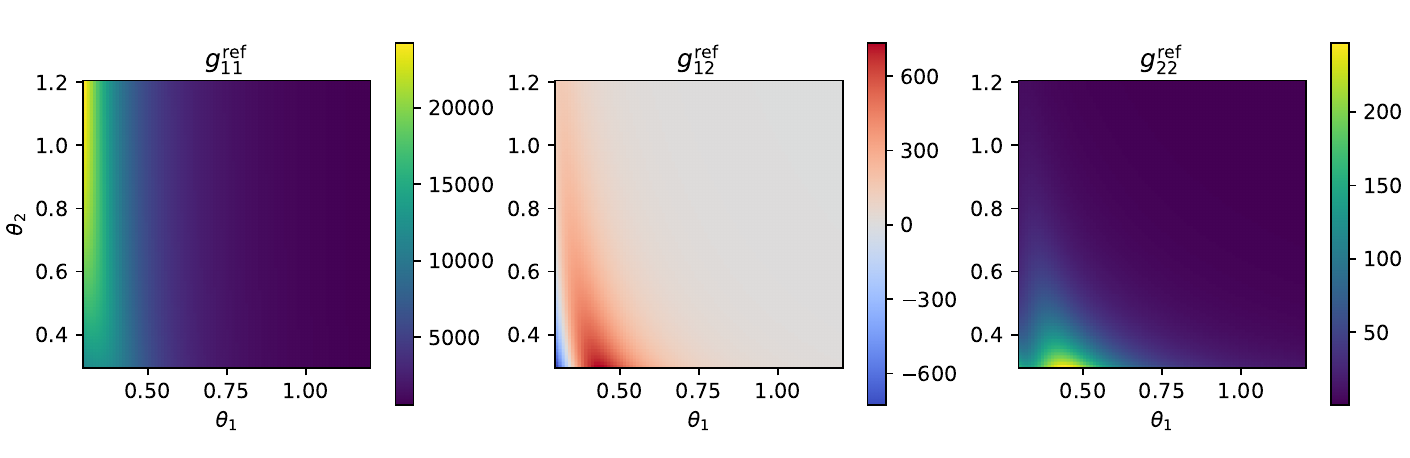}
 \caption{Components of the reference thermodynamic metric for a
 normalized reflecting harmonic equilibrium and the interior-wedge
 reference resistance. Here \(a=200\) and \(\Gamma=k_0/(\eta b)\).}
 \label{fig:scallop-metric-ref}
\end{figure}

\begin{figure}[t]
 \centering
 \includegraphics[width=.96\linewidth]{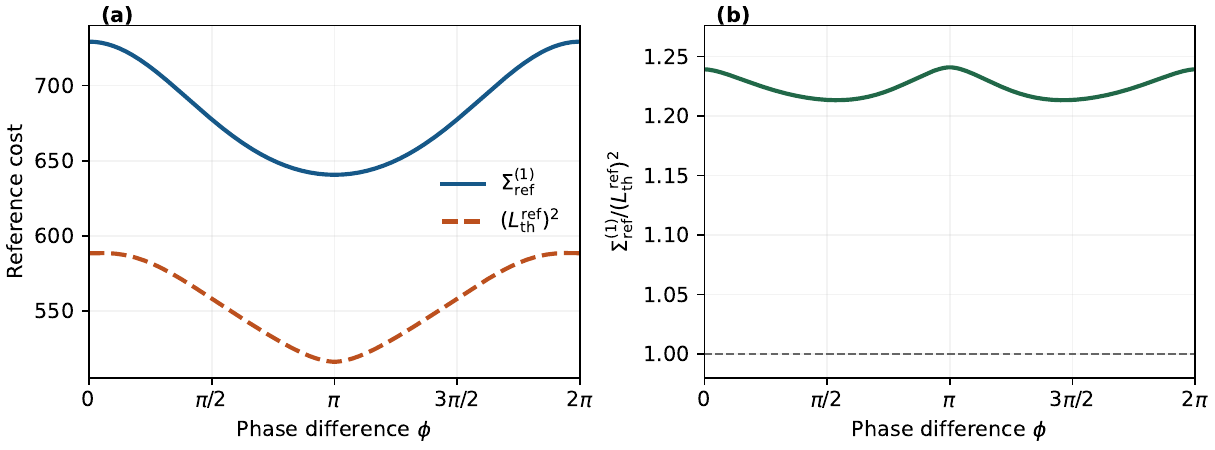}
 \caption{Thermodynamic-length bound for the \(1:1\) sinusoidal protocol
 with the reference metric of Fig.~\ref{fig:scallop-metric-ref}.
 (a) Leading entropy-production coefficient \(\Sigma_{\rm ref}^{(1)}\)
 and squared length \((\mathcal L_{\rm th}^{\rm ref})^2\) versus \(\phi\).
 (b) Their ratio; the line denotes the constant-speed bound.}
 \label{fig:scallop-length-ref}
\end{figure}

Figure~\ref{fig:scallop-metric-ref} shows the three independent components of the reference metric. 
We set $a=200$ and $\Gamma=k_0/(\eta b)$, where $\eta$ is the fluid viscosity, and $b$ is the fixed out-of-plane span of the reference wedge (Appendix~\ref{app:hydrodynamic_reduction}), and use $\pi/18\le x\le\pi$.
For Fig.~\ref{fig:scallop-length-ref}, we use the \(1:1\) protocol
\begin{align}
 \theta_1(s)&=0.65+0.11\sin(2\pi s),\\
 \theta_2(s)&=0.65+0.22\sin(2\pi s+\phi),\qquad 0\le s\le1.
\end{align}
Changing \(\phi\) changes the path, not just its time parametrization.
Writing \(v_{\rm th}^2=\theta_i' g^{\rm ref}_{ij}\theta_j'\), we compute
\begin{equation}
 \Sigma_{\rm ref}^{(1)}=\int_0^1 v_{\rm th}^2\,ds,\quad
 \mathcal L_{\rm th}^{\rm ref}=\int_0^1 v_{\rm th}\,ds,\quad
 R_{\rm th}=\frac{\Sigma_{\rm ref}^{(1)}}{(\mathcal L_{\rm th}^{\rm ref})^2}.
\end{equation}
The leading cycle entropy production in this reference model is
\(\delta\Sigma_{\rm ref}^{(1)}\), not \(\Sigma_{\rm ref}^{(1)}\).
The ratio satisfies \(R_{\rm th}\ge1\), with equality if the given
path is traversed at constant thermodynamic speed. It compares the
uniform-phase schedule with the constant-speed dissipation bound for each path; it is not
a propulsion efficiency. In particular, \(\phi=0,\pi\) gives a
reciprocal path with zero geometric displacement but finite dissipation.

\subsection{A passive-slip hinge model and a testable stroke}
\label{sec:passive-slip-hinge}

To exhibit a boundary control that changes the force-free connection
without changing the body geometry, consider two mirror-symmetric
hinged arms in a Newtonian fluid. Each arm carries two fixed-radius
spheres of radius $R$, at distances $\ell_p$ and $\ell_d$ from the
hinge, with $\ell_p<\ell_d$. We take all sphere separations to
be large compared with $R$ throughout the stroke and treat the
supporting links and hinge as hydrodynamically negligible.
In this leading independent-sphere approximation, the proximal
spheres are no-slip, with drag coefficient $c_p=6\pi\eta R$.
The distal spheres share a uniform, passive Navier slip length
$\lambda$ and have translational drag
\begin{equation}
c_d(\lambda)=6\pi\eta R\,
 \frac{1+2\lambda/R}{1+3\lambda/R}.
\label{eq:distal-slip-drag}
\end{equation}
See, for example, the steady term in Eq.~(5) of
Ref.~\cite{Legendre2019}. The boundary control changes the slip
coefficient, not the prescribed surface velocity; it produces
no flow by itself for a stationary swimmer in this idealization.

Taking $X$ positive opposite to the direction toward the tips,
the axial velocity of a bead at distance $\ell_j$ from the hinge
is $U+\frac12\ell_j\sin(x/2)\dot x$. The total axial force is the
sum of the four Stokes drags. Force balance therefore gives
\begin{align}
U&=-G_{\rm bead}(x,\lambda)\dot x,\\
G_{\rm bead}(x,\lambda)
&=\frac{\sin(x/2)}{2}\,
  \frac{c_p\ell_p+c_d(\lambda)\ell_d}{c_p+c_d(\lambda)}.
\label{eq:bead-connection}
\end{align}
Thus changing the distal drag relative to the proximal drag
changes the drag-weighted lever arm. Multiplying all drag
coefficients by a common factor leaves this lever arm unchanged.
The angle dependence $\sin(x/2)$ and the slip-dependent lever arm
factorize, so the same argument leading to
Eq.~\eqref{eq:finite-rate-displacement} applies with a primitive
and modulation calibrated for this swimmer. This model does
not derive Eq.~\eqref{G(x,theta_2)} for the Kim plates.

Consider a prescribed stroke that opens from $x_a$ to $x_b$
in one slip state and closes over the same angular interval
in the other. Switch the surface state while holding the hinge
at the turning angles. The predicted signed axial displacement is
\begin{equation}
\Delta X=\int_{x_a}^{x_b}
 \bigl[G_{\rm bead}(x,\lambda_{\rm close})
       -G_{\rm bead}(x,\lambda_{\rm open})\bigr]\,dx .
\label{eq:bead-stroke}
\end{equation}
For illustration, let $\ell_p=L/3$, $\ell_d=L$,
$x_a=0.40$, and $x_b=0.90$ (in radians). Taking
$\lambda_{\rm open}=R$ and $\lambda_{\rm close}=0$ gives
$\Delta X/L=[\cos(x_a/2)-\cos(x_b/2)]/21
\simeq 3.79\times10^{-3}$ at leading order.
This is a model estimate for a specified slip contrast, not a
measured displacement or a claim that such a switch has been
implemented. Finite bead separations and support drag would
modify the estimate.

\section{Discussion}
\label{sec:discussion}

The cycle displacement separates a kinematic condition from the
statistics of the opening angle. If the connection depends only on
$x$, the shape contribution to translation is an exact differential.
Its mean over a cycle is therefore zero whenever the shape
distribution returns to itself, including in a periodic state at
finite driving rate. This conclusion allows for translational
diffusion; it rules out a mean displacement, not displacement
fluctuations. In the present model an independently controlled
hydrodynamic response removes the exact-differential restriction.
It supplies a second control without adding a second shape
coordinate. This is outside the shape-only kinematics to which
Purcell's scallop theorem applies.

The deterministic limit clarifies what the thermal fluctuations do.
For the protocol studied here, whose equilibrium angles stay in the
interior of the reflecting interval, the curvature tends to
$(\epsilon/\bar\theta_2)G_{\rm Kim}(\theta_1)$ as $T\to0$,
when the slow-driving limit is taken first. Both the curvature
and the computed cycle displacement can thus remain nonzero without
thermal fluctuations. For the selected circular stroke, broadening
the angle distribution reduces the displacement over the
temperature range displayed in Fig.~\ref{fig:scallop-temp-ref}.
This trend need not hold for other connections or protocols. The
large-$T$ decay in the present model has a different and more general
explanation: on the fixed reflecting interval, the equilibrium
density becomes uniform and its sensitivity to the controls
vanishes.

The most direct test of the mechanism in
Sec.~\ref{sec:passive-slip-hinge} is to calibrate
$G=-U/\dot x$ over the same prescribed angular stroke at two
fixed boundary settings, with the swimmer aligned but free to
translate axially. Integrating the difference of the two
calibration curves predicts the signed displacement for the
switched stroke. Interchanging the settings on opening and
closing should reverse the sign, whereas using the same setting
on both legs should give zero cycle displacement. A separate
test with the hinge locked should monitor transient flow
and axial drift for each switch at both turning angles, to
check for flow driven directly by switching. The key experimental
challenge is to obtain a measurable drag contrast without such
direct forcing. Once a factorized connection is calibrated,
Eq.~\eqref{eq:finite-rate-displacement} provides a further test
for fluctuating strokes: the measured shape moment and the
measured displacement can be compared over a range of driving
rates.

The thermodynamic metric addresses a different question. For a
prescribed path and positive mobility, the thermodynamic-length
bound gives the lowest leading shape-sector dissipation over time
parametrizations at fixed period. Constant thermodynamic speed
attains this bound; choosing a different path is a separate
optimization that must also account for the signed curvature flux.
The interior-wedge mobility in
Figs.~\ref{fig:scallop-metric-ref} and \ref{fig:scallop-length-ref}
illustrates the length bound but does not provide a calibrated
clamped resistance or a bound on the total energy cost of swimming.
The exact clamped versus force-free comparison is conditional on
both mobilities being derived from the same positive-definite
resistance matrix. In the reference example, reciprocal paths
can have zero geometric displacement and nonzero dissipation.

\section{Concluding Remarks}
\label{sec:conclusion}

We have separated the mean cycle displacement of a stochastic
single-hinge scallop from the leading dissipative cost of driving
its shape. In the slow-driving regime the former is a flux of
geometric curvature, while the latter is described by a
thermodynamic metric. For a shape-only hydrodynamic connection,
the mean displacement vanishes in a periodic state even outside
the slow-driving regime. A second control that modulates the
connection permits a nonzero curvature, and the harmonic example
shows that the resulting displacement can survive in the
deterministic limit. Thermal fluctuations change the sampling
of the connection but are not the source of the effect.

For the factorized connection, the periodic-state displacement
can be obtained from the time-dependent angle distribution
at any driving rate. The passive-slip hinge model turns the
hydrodynamic mechanism into a prediction for a reciprocal
angular stroke with different boundary settings on its two legs.
An experimental implementation would need to calibrate the
change in the axial connection and check for direct flow from
the boundary actuator. The thermodynamic metric used here does
not include the operating cost of that actuator.

\begin{acknowledgments}

The author thanks Jorge Kurchan for his suggestion of this problem. 
This work was partially supported by the JSPS KAKENHI Grant No. JP26K06960.

\end{acknowledgments}

\vspace{1cm}
\appendix

\section{Hydrodynamic Reference Models}
\label{app:hydrodynamic_reduction}

The shape resistance and the translation--shape connection play
different roles. Below we derive a shape resistance from the
dissipation in an interior wedge and take the force-free connection
from a separate finite-plate calculation. These two ingredients are
not the entries of a single calibrated resistance matrix.

\subsection{Interior-wedge reference resistance}
\label{subsec:app_rotational_resistance}

Consider two plates of length $L$ at $\vartheta=\pm x/2$. We use the
interior similarity flow of Moffatt~\cite{Moffatt1964}, truncated at
$r=L$. Set
\begin{align}
 D(x)&=\sin x-x\cos x,\nonumber\\
 f(\vartheta;x)&=
 \frac{\sin(2\vartheta)-2\vartheta\cos x}{D(x)},\nonumber\\
 \Psi(r,\vartheta)&=-\frac{\dot x}{4}r^2 f(\vartheta;x).
 \label{eq:app_Moffatt_f2}
\end{align}
With $u_r=r^{-1}\partial_\vartheta\Psi$ and
$u_\vartheta=-\partial_r\Psi$, the boundary velocities are
$u_r(r,\pm x/2)=0$ and
$u_\vartheta(r,\pm x/2)=\pm r\dot x/2$. The shear traction parallel to
a radial plate produces no moment about the hinge; its magnitude
cannot be integrated with a moment arm to obtain the rotational
resistance. Instead, we integrate the viscous strain-rate dissipation
in the interior fluid. In this flow the polar strain components are
$E_{rr}=-\dot x f'/4$, $E_{\vartheta\vartheta}=\dot x f'/4$, and
$E_{r\vartheta}=-\dot x f''/8$, with primes denoting angular
derivatives. Thus the dissipated power per unit span is
\begin{align}
 P_{\mathrm{wedge}}&=\int_0^L r\,dr
 \int_{-x/2}^{x/2}2\eta E:E\,d\vartheta
 =\zeta_{\rm ref}^{\rm(per,span)}(x)\dot x^{\,2},\nonumber\\
 \zeta_{\rm ref}^{\rm(per,span)}(x)
 &:=\frac{\eta L^2}{32}\int_{-x/2}^{x/2}
       \big[4(f')^2+(f'')^2\big]d\vartheta=\frac{\eta L^2}{2}
 \frac{x(1+\cos^2 x)-2\sin x\cos x}{(\sin x-x\cos x)^2}.
 \label{eq:scallop-ref-resistance}
\end{align}
In particular, $\zeta_{\rm ref}^{\rm(per,span)}\sim3\eta L^2/(2x^3)$ for a narrow
gap and $\zeta_{\rm ref}^{\rm(per,span)}(\pi)=\eta L^2/\pi$. This construction
excludes the exterior fluid and does not impose the finite-plate end
condition at $r=L$. It provides a positive angle-dependent reference
resistance, not the exact clamped resistance of finite plates or a
rigorous bound on it.

Let $b$ be the fixed out-of-plane span assigned to the reference
wedge. The rotational resistance used in the stochastic model is
$\zeta_{\rm ref}(x)=b\,\zeta_{\rm ref}^{\rm (per\,span)}(x)$.
Here $\eta$ is the ordinary dynamic viscosity. Multiplying by $b$
converts the power per unit span to total reference power; it does
not account for flow near the spanwise ends.

\subsection{Force-free connection and the Kim data}
\label{subsec:app_cross_resistance}

Kim, Jeong, and Lee~\cite{Kim1986} consider two plates at
$\vartheta=\pm\alpha$ rotating with angular velocities $\mp\Omega$,
so $\dot x=2\dot\alpha=-2\Omega$. Their $U_\infty$ is the far-fluid
velocity in the hinge-fixed frame, with the positive axial direction
pointing toward the tips. The hinge velocity in a fluid at rest is
therefore $-U_\infty$ along the Kim axis. We choose positive $X$
opposite to that axis; in this convention
$\dot X=-G_{\mathrm{Kim}}(x)\dot x$ for the baseline connection, with
$G_{\mathrm{Kim}}/L=U_\infty/(2\Omega L)$ at $\alpha=x/2$,
as in Eq.~\eqref{eq:G_Kim}. Replacing the hinge by the center of mass
of the plates adds only an exact shape differential to $dX$ and
does not change the cycle displacement in a periodic state.

For the numerical demonstration, we use a recalibrated approximate
digitization of the upper $U_\infty/(\Omega L)$ branch in Fig.~4 of
Ref.~\cite{Kim1986}. The 24 points span $5^\circ\le\alpha\le90^\circ$
and include the exact endpoint $G_{\mathrm{Kim}}(\pi)/L=1/\pi$.
We use piecewise-linear interpolation in $x=2\alpha$ and do not
extrapolate toward the closed hinge. The data give only the baseline
angular dependence; the stiffness modulation in
Eq.~\eqref{G(x,theta_2)} is phenomenological.

\begin{table}[tb]
\centering
\caption{Interpolation knots used for the baseline connection in
Figs.~\ref{fig:scallop-curv-ref}--\ref{fig:scallop-area-ref}.
The half-opening angle $\alpha$ is given in degrees, so
$x=\pi\alpha/90$ is the full opening angle in radians.
For $\alpha<90^\circ$ the values are approximate readings of the
upper solid branch in Fig.~4 of Kim \textit{et al.}~\cite{Kim1986},
converted using $G_{\rm Kim}/L=U_\infty/(2\Omega L)$.
The endpoint at $90^\circ$ uses the exact flat-plate value.
The displayed digits specify the interpolation inputs, not the
accuracy of the digitization. We interpolate linearly in $x$ and
do not extrapolate.}
\label{tab:kim-fig4-knots}
\begin{tabular}{rr@{\qquad\qquad}rr}
\hline\hline
$\alpha$ (deg) & $G_{\rm Kim}/L$ & $\alpha$ (deg) & $G_{\rm Kim}/L$ \\
\hline
5 & 0.60735 & 35 & 0.33470 \\
6 & 0.56235 & 40 & 0.32855 \\
7 & 0.52825 & 45 & 0.32515 \\
8 & 0.50070 & 50 & 0.32270 \\
9 & 0.47950 & 55 & 0.32165 \\
10 & 0.46155 & 60 & 0.32120 \\
12 & 0.43265 & 65 & 0.32045 \\
15 & 0.40570 & 70 & 0.32035 \\
18 & 0.38670 & 75 & 0.31990 \\
20 & 0.37590 & 80 & 0.31950 \\
25 & 0.35715 & 85 & 0.31875 \\
30 & 0.34430 & 90 & $1/\pi$ \\
\hline\hline
\end{tabular}
\end{table}

\subsection{Clamped versus force-free mobility}
\label{app:mobility-derivation}

For the same symmetric positive-definite resistance matrix in
Eq.~\eqref{eq:resistance_matrix}, the internal resistance with
$\dot X=0$ is $\zeta_{xx}$, whereas that with
$F_{\mathrm{prop}}=0$ is its Schur complement
$\zeta_{xx}-\zeta_{Xx}^2/\zeta_{XX}$. Inverting these positive
resistances gives $M_{\mathrm{eff}}\geq M_{\mathrm{cl}}$, with equality
when the cross-resistance vanishes. This comparison applies
pointwise at fixed shape and controls, when the resistance matrix
is well defined, as in a confined or genuinely three-dimensional
system. We do not use the unbounded two-dimensional calculation
of Ref.~\cite{Kim1986} to infer an externally forced translational
resistance or to identify the reference resistance in
Eq.~\eqref{eq:scallop-ref-resistance} with $\zeta_{xx}$.

\section{Slow-Driving Expansion and Generalized Inverse}
\label{app:slow-driving}

\subsection{Expansion at fixed reduced protocol}
\label{app:slow_driving_details}

For the periodic state with a smooth prescribed protocol
$\bm\theta(s)$, use $s=t/\tau_p$ and $\delta=(\Gamma\tau_p)^{-1}$ as in
Eq.~\eqref{eq:delta_s}. A formal expansion
$\rho=\rho^{\mathrm{ss}}+\sum_{n\geq1}\delta^n\rho^{(n)}$
has $\int_\Omega\rho^{(n)}dx=0$ for $n\geq1$. The recurrence obtained
from Eq.~\eqref{eq:dimensionless_smoluchowski} is
\begin{equation}
 \widetilde{\mathcal L}_{\bm\theta(s)}\rho^{(n)}(x,s)
 =\partial_s\rho^{(n-1)}(x,s),\qquad
 \rho^{(n)}=\widetilde{\mathcal L}_{\bm\theta(s)}^+
                \partial_s\rho^{(n-1)},
 \label{eq:app_pn_solution}
\end{equation}
with $\rho^{(0)}=\rho^{\mathrm{ss}}$. The right-hand side has zero
integral because the boundaries are fixed and the previous term
has constant normalization. At higher orders, $\partial_s$ acts
also on the instantaneous generalized inverse where it occurs
in $\rho^{(n-1)}$. The slow-driving limit is taken with the protocol
held fixed; its period must be long relative to the relevant
relaxation times along the path.

\subsection{Reversibility and generalized inverse}
\label{app:generalized-inverse}

At any fixed controls, Eq.~\eqref{eq:dimensionless_generator} has the
reversible form
\begin{equation}
 \widetilde{\mathcal L}_{\bm\theta}r
 =\partial_x\!\left[
 \widetilde M\rho^{\mathrm{ss}}
 \partial_x(r/\rho^{\mathrm{ss}})\right].
 \label{eq:app_reversible_generator}
\end{equation}
With reflecting boundary conditions, integration by parts gives
\begin{equation}
 \int_\Omega\frac{r^*
       (-\widetilde{\mathcal L}_{\bm\theta}r)}{\rho^{\mathrm{ss}}}\,dx
 =\int_\Omega\widetilde M\rho^{\mathrm{ss}}
       \left|\partial_x(r/\rho^{\mathrm{ss}})\right|^2dx.
 \label{eq:app_dirichlet_form}
\end{equation}
The operator is symmetric in the weighted density inner product
(with weight $1/\rho^{\mathrm{ss}}$). On a regular finite interval with
strictly positive mobility, its self-adjoint realization has a
stationary mode $\rho^{\mathrm{ss}}$ and a gapped relaxation spectrum.
The left stationary mode in the ordinary density pairing is $1$.
The generalized inverse in Eq.~\eqref{eq:pseudoinverse_spectral}
satisfies
\begin{align}
 \widetilde{\mathcal L}^+\widetilde{\mathcal L}
 &=\widetilde{\mathcal L}\widetilde{\mathcal L}^+
 =I-|\rho^{\mathrm{ss}}\rangle\langle 1|,
 \label{eq:app_c1}\\
 \widetilde{\mathcal L}^+\rho^{\mathrm{ss}}&=0,\qquad
 \int_\Omega (\widetilde{\mathcal L}^+r)(x)\,dx=0.
 \label{eq:app_int_zero}
\end{align}
Thus the inverse selects a unique zero-integral correction.

Independently, twice differentiating
$\int_\Omega\rho^{\mathrm{ss}}dx=1$ at fixed boundaries yields the
equilibrium Fisher identity
\begin{align}
 -\int_\Omega \rho^{\mathrm{ss}}\partial_i\partial_j
 \ln\rho^{\mathrm{ss}}\,dx
 &=\int_\Omega
 \frac{(\partial_i\rho^{\mathrm{ss}})(\partial_j\rho^{\mathrm{ss}})}
 {\rho^{\mathrm{ss}}}\,dx .
 \label{eq:Fisher=Hessian_cont}
\end{align}
This identity does not involve the generalized inverse and should
not be identified with the dissipative metric.

\section{Full-Line Constant-Mobility Benchmark}
\label{app:constant_mobility_benchmark}

The full-line harmonic model with constant mobility $M_0=1/\gamma$
provides a check on the metric in Eq.~\eqref{eq:metric_tensor}.
Unlike the numerical reflecting model, this benchmark has no
finite endpoints. Its dimensionless mobility is
$\widetilde M_0=T/(\Gamma\gamma)$.
Writing $\sigma^2=T/(k_0L^2\theta_2)$, its stationary density is
$\rho^{\mathrm{ss}}=(\sqrt{2\pi}\sigma)^{-1}
 e^{-(x-\theta_1)^2/(2\sigma^2)}$. Its cumulative derivatives are
$S_1=-\rho^{\mathrm{ss}}$ and
$S_2=(x-\theta_1)\rho^{\mathrm{ss}}/(2\theta_2)$.
Gaussian moments then give
\begin{equation}
 g_{11}^{(0)}=\frac{1}{\widetilde M_0},\qquad
 g_{12}^{(0)}=g_{21}^{(0)}=0,\qquad
 g_{22}^{(0)}
 =\frac{T}{4 k_0 L^2\theta_2^3\widetilde M_0}.
 \label{eq:gij_constant_dimensionless}
\end{equation}
Here $(0)$ labels the constant-mobility model, not a
slow-driving order. The cross term vanishes by odd parity about
$\theta_1$. On a finite interval it need not vanish even for a
constant mobility. The resulting first-order coefficient is
\begin{equation}
 \Sigma_0^{(1)}=
 \frac{1}{\widetilde M_0}\int_0^1
 \left[(\theta_1')^2+
 \frac{T}{4 k_0 L^2\theta_2^3}(\theta_2')^2\right]ds.
 \label{eq:Sigma1_constant_benchmark}
\end{equation}
In Eq.~\eqref{eq:Sigma_ex} the leading physical entropy production is
$\delta\Sigma_0^{(1)}$ for this benchmark.

\end{document}